\documentclass[aps,prb,amsmath,twocolumn,floatfix]{revtex4}
\usepackage{color}
\usepackage{graphicx}
\usepackage{amssymb, amsmath, amsfonts}
\begin{document}

\title{Controlling the Colour of Metals: Intaglio and Bas-Relief Metamaterials}

\author{Jianfa Zhang$^1$}
\author{Jun-Yu Ou$^1$}
\author{Nikitas Papasimakis$^1$}
\author{Yifang Chen$^2$}
\author{Kevin F. MacDonald$^1$}
\email[]{kfm@orc.soton.ac.uk}
\homepage[]{www.nanophotonics.org.uk}
\author{Nikolay I. Zheludev$^1$}
\affiliation{$^1$Optoelectronics Research Centre \& Centre for Photonic Metamaterials, University of Southampton, Highfield, Southampton, Hampshire, SO17 1BJ, UK
\\
$^2$Rutherford Appleton Laboratory, Harwell Science and Innovation Campus, Didcot, Oxfordshire, OX11 0QX, UK}

\date{\today}

\begin{abstract}
\noindent
The fabrication of indented (`intaglio') or raised (`bas-relief') sub-wavelength metamaterial patterns on a metal surface provides a mechanism for changing and controlling the colour of the metal without employing any form of chemical surface modification, thin-film coating or diffraction effects. We show that a broad range of colours can be achieved by varying the structural parameters of metamaterial designs to tune absorption resonances. This novel approach to the `structural colouring' of pure metals offers great versatility and scalability for both aesthetic (e.g. jewellery design) and functional (e.g. sensors, optical modulators) applications. We focus here on visible colour but the concept can equally be applied to the engineering of metallic spectral response in other electromagnetic domains.
\end{abstract}

\maketitle
\noindent

Human visual perception of surface colour is determined by its properties of reflection, transmission, and absorption at wavelengths between about 400 and 700~nm. Most pure metals are essentially colourless because, with plasma frequencies in the ultraviolet domain, their valence electrons are able to absorb and re-emit photons efficiently over the entire visible spectral range. Gold and copper - obvious exceptions to this rule - have lower plasma frequencies and consequently absorb blue and blue/green light respectively, thereby achieving their characteristic yellow and red/orange colours\cite{Tilley1999}. This limited palette of metallic colours is typically extended or replaced through the application of coatings (such as paint and multilayer dielectrics) or controlled chemical modification (e.g. oxide formation by anodization). In the natural world, many plants and animals display dramatic `structural colours' derived from astonishingly intricate three-dimensional assemblies of intrinsically colourless bio-materials\cite{Parker2000}. While the physics of these colours is in many cases well understood, replicating them remains a significant challenge\cite{Parker2007,Kolle2010,Huang2006} and typically requires complex (multi- or atomic) layer deposition and etching fabrication procedures.

Here, we report on a form of structural colouring for pure metals that relies only on the formation of arrays of sub-wavelength elements inscribed into or raised above the surface to a depth/height of the order 100~nm. These intaglio and bas-relief patterns constitute a new family of planar metamaterials: Conventional forms comprise discrete plasmonic `meta-molecules' distributed on a dielectric substrate or meta-molecule voids perforating a metallic thin film\cite{Luk'yanchuk2010}. As such they present a discontinuous metallic structure to incident electromagnetic radiation. In contrast, the meta-surfaces considered here (wherein a patterned `layer' effectively sits on a `substrate' of the same metal, as shown in Fig.~1a) present a continuous metallic profile to incident light. These designs can be engineered to selectively provide highly efficient resonant absorption, thereby dramatically changing the metal's reflection spectrum and perceived colour (Fig.~1b). The colours produced can, by design, be polarization-dependent or -independent and are relatively insensitive to viewing angle.

\begin{figure}
\includegraphics[width=80mm]{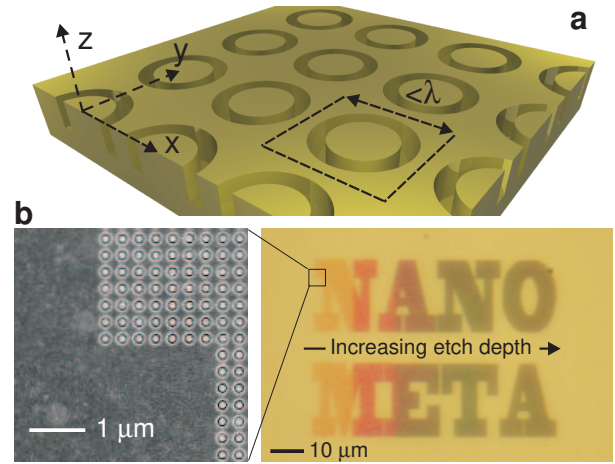}
\caption{\textbf{Metallic structural colour.} \textbf{a}, Artistic impression of an intaglio metamaterial array of sub-wavelength single ring meta-molecules inscribed into a metal surface. \textbf{b}, The realization of this concept in gold: The words `NANO META' seen under an optical microscope on the right are formed from arrays of 170~nm diameter rings (as shown in the electron microscope image, left) milled to a depth that increases in six steps from 60 to 200~nm across the sample.$^{\dag}$}
\end{figure}

\begin{figure}[h!]
\includegraphics[width=80mm]{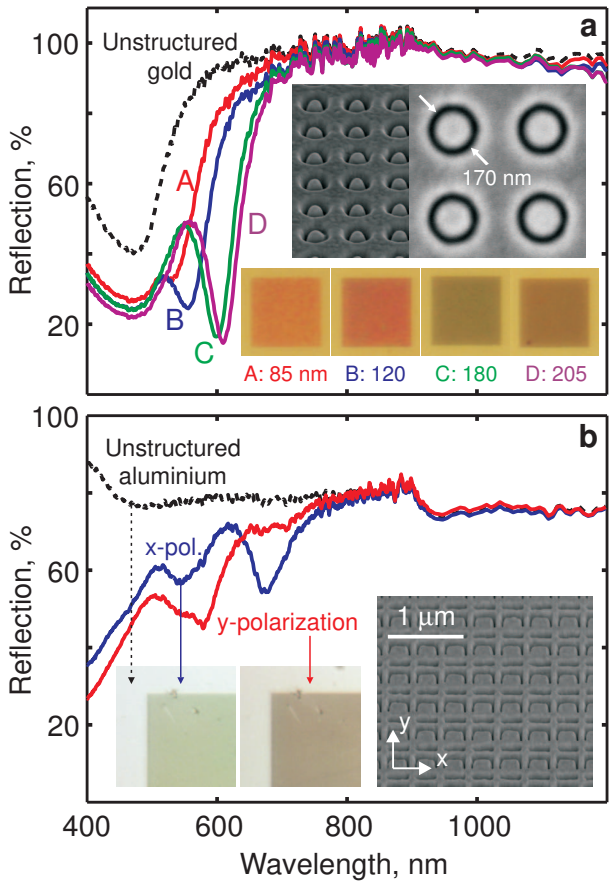}
\caption{\textbf{Changing the colour of gold and aluminium.} \textbf{a}, Reflection spectra for an unstructured gold surface and for the same surface patterned with 170~nm diameter rings cut to depths ranging from 85 to 205~nm (as labelled). The insets show electron microscope images of the intaglio metamaterial design (at oblique incidence and in plan view) and optical microscope images of the different patterned domains. \textbf{b}, Reflection spectra for an unstructured aluminium/silica interface and for regions of the same sample patterned with a raised (bas-relief) metallic pattern of asymmetric split rings for incident polarizations parallel and perpendicular to the split. The insets show an electron microscope image of the split ring design etched in silica prior to aluminium deposition and optical microscope images of the patterned domain for the two orthogonal incident light polarizations.$^{\dag}$}
\end{figure}

A microspectrophotometer was employed to measure normal incidence reflection spectra for a variety of gold and aluminium metamaterial designs. Gold intaglio metamaterial patterns were fabricated by focused ion beam milling on 250~nm thick gold films evaporated on glass substrates. Fig.~2a shows spectra for square arrays of 170~nm single rings milled to four different depths into a gold film, alongside the reflection spectrum for the adjacent unstructured gold surface. The red shift of absorption resonance with increasing etch depth is clearly seen and the associated changes in the colour of gold are strikingly illustrated by the inset optical microscope image. Aluminium bas-relief structures were fabricated at an interface between the metal and an optically polished fused silica substrate using electron beam lithography and anisotropic reactive ion etching: Patterns were etched into the silica to a nominal depth of 70~nm then coated by evaporation with a 250~nm layer of aluminium. Fig.~2b shows spectra for an anisotropic aluminium bas-relief design of asymmetric split rings. Again the change in colour from that of the unstructured metal is dramatic, and in this case the effect is, by design, dependent on the polarization of incident light.

\begin{figure}
\includegraphics[width=80mm]{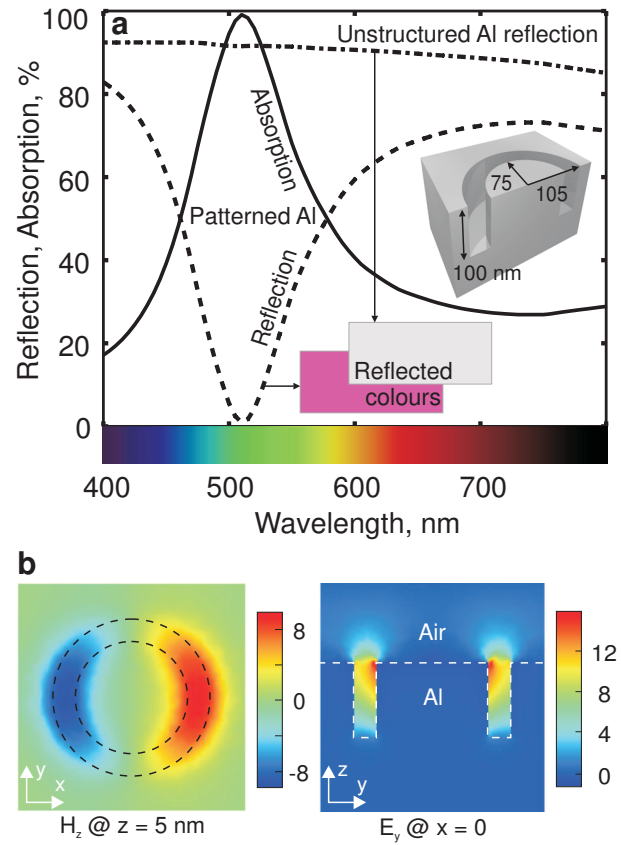}
\caption{\textbf{Colour by design.} \textbf{a}, Numerically simulated reflection and absorption spectra for an aluminium surface patterned with a square array 100~nm deep intaglio metamaterial design of 210~nm single rings alongside the reflection spectrum for unstructured aluminium and inset blocks showing the perceived colours of these surfaces. \textbf{b}, Maps of surface-normal magnetic field intensity in a z-plane 5~nm above the top surface of the metal (left) and electric field intensity parallel to the metal surface in an x-plane that diametrically bisects a ring (right). The structure is illuminated at normal incidence with y-polarized plane waves.$^{\dag}$}
\end{figure}

The sub-wavelength periodicity of the metamaterials designs preclude diffraction effects and the colours achieved by metallic surfaces structured in this way are related to the plasmonic resonances of the meta-molecules. Fig.~3 presents a numerical analysis of a single-ring aluminium intaglio design: the unstructured metal uniformly reflects more than 80\% of light across the range from 400 to 800~nm and is consequently seen to be light grey in colour (Fig.~3a). An array of 210~nm rings milled to a depth of 100~nm below the surface endows the metal with a strong absorption resonance centred in the green part of the spectrum at 510~nm. The associated substantial modification of its reflection spectrum gives the aluminium a vivid magenta colour. The electromagnetic field maps shown in Fig.~3b illustrate how this absorption is linked, in this particular case, to a circulating slot mode of the ring structure\cite{Waele2009} with azimuthal mode number $m$=1.

By adjusting meta-molecule geometries one can achieve a wide palette of colours. Fig.~4 illustrates how variations in the simple single ring intaglio design on aluminium and gold can provide access to a significant proportion of colour space. The parameter space for meta-molecule design and distribution is obviously almost unlimited and different geometries will undoubtedly extend the accessible colour range. That said, the `pure' colours found at the boundaries of the CIE1931 standard colour space presented in Fig.~4 are likely to present a significant challenge (they would require suppression of reflectivity across all but a narrow band of the visible range).

\begin{figure}
\includegraphics[width=80mm]{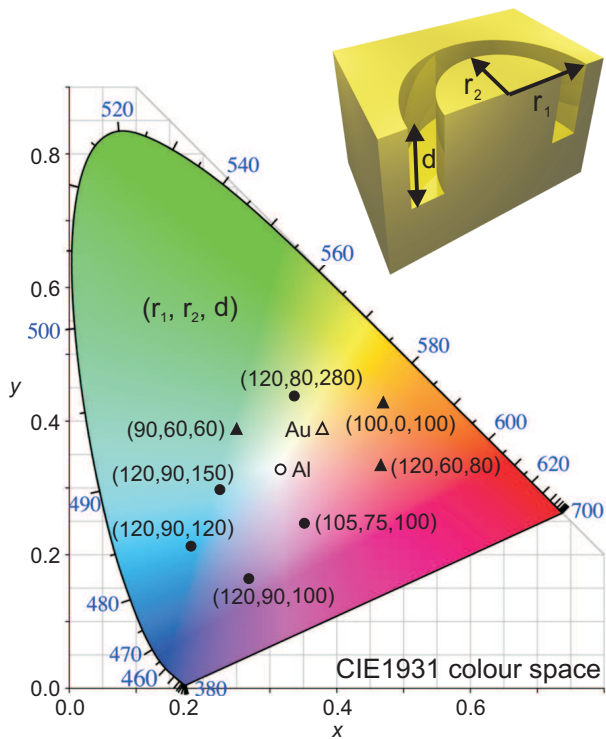}
\caption{\textbf{Accessible metallic colours.} \textbf{a}, CIE1931 chromaticity diagram overlaid with points corresponding to the simulated reflected colours of single ring intaglio metamaterial designs in aluminium (circular symbols) and gold (triangles), labelled according to their structural parameters (external radius~$r_1$, internal radius~$r_2$, depth~$d$, array period =~300~nm in all cases). Points for the unstructured metals are indicated by open symbols.$^{\dag}$}
\end{figure}

The intaglio/bas-relief metamaterial approach provides a mechanism for dramatically changing and controlling the colour, or more broadly the spectral response, of metal surfaces whilst retaining other properties (e.g. smoothness, hardness, lustre, electrical conductivity) that would be lost in the use of materially different coatings. In being composed of `pattern' and `substrate' layers of the same medium, such structures offer considerable advantages in ease of fabrication and application to bulk (as opposed to thin-film) media and/or non-planar surface profiles. They may be produced with minimal adaptations to standard metal-forming process (e.g. pressing, rolling, casting) and applied to anything from an item of jewellery to a piece of automotive bodywork. At the same time, such structures may provide optical properties that are extremely difficult to imitate, thereby facilitating security applications (e.g. banknote anti-forgery features) and providing high-value exclusivity in aesthetic applications.\\

\noindent
\textbf{Acknowledgements:} This work was supported by the UK's Engineering and Physical Sciences Research Council (all authors), The Royal Society (NIZ) and the China Scholarship Council (JZ).\\

\noindent
$^{\dag}$ Colour presentation: The balance of optical microscope images is referenced to a test pattern of 24 RGB-specified sample colours (ColourChecker Mini by X-Rite). Perceived colours are derived from experimentally measured and numerically simulated reflection spectra using Judd-Vos-modified CIE 2-deg colour matching functions\cite{Vos1978,Wyszecki1982,CVRL} assuming an illuminating light source with the spectral radiant power distribution of a 6500~K black body. Readers should be aware that their monitor and/or printer settings may affect colour rendering (Fig.~5 shows reference colours with specified RGB values for comparison).

\begin{figure}[h!]
\includegraphics[width=78mm]{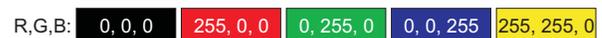}
\caption{\textbf{Reference colours.}}
\end{figure}

\end{document}